\documentclass{article}

\usepackage{microtype}
\usepackage{graphicx}
\usepackage{subfigure}
\usepackage{booktabs} 

\usepackage{amsmath} 
\usepackage{amsthm,amssymb,amsfonts}
\graphicspath{{figures/}} 
\usepackage{makecell}
\usepackage[thinlines]{easytable}
\usepackage{adjustbox}

\usepackage{hyperref}

\usepackage[accepted]{icml2020}

\icmltitlerunning{SimGANs: Simulator-Based Generative Adversarial Networks for ECG Synthesis to Improve Deep ECG Classification}

\newcommand{\SimDCGAN}{SimDCGAN}
\newcommand{\SimVGAN}{SimVGAN}
\newcommand{\SimGAN}{SimGAN}
\newcommand{\VGAN}{VGAN}

\begin{document}

\twocolumn[
\icmltitle{SimGANs: Simulator-Based Generative Adversarial Networks \\
           for ECG Synthesis to Improve Deep ECG Classification}




\begin{icmlauthorlist}
\icmlauthor{Tomer Golany}{technion}
\icmlauthor{Daniel Freedman}{google}
\icmlauthor{Kira Radinsky}{technion}
\end{icmlauthorlist}

\icmlaffiliation{technion}{Technion - Israel Institute of Technology, Haifa, Israel}
\icmlaffiliation{google}{Google Research}

\icmlcorrespondingauthor{Tomer Golany}{tomer.golany@cs.technion.ac.il}
\icmlcorrespondingauthor{Kira Radinsky}{kirar@cs.technion.ac.il}
\icmlcorrespondingauthor{Daniel Freedman}{danielfreedman@google.com}

\icmlkeywords{Machine Learning, ICML}

\vskip 0.3in
]



\printAffiliationsAndNotice{}  

\begin{abstract}
Generating training examples for supervised tasks is a long sought after goal in AI. We study the problem of heart signal electrocardiogram (ECG) synthesis for improved heartbeat classification. ECG synthesis is challenging: the generation of training examples for such biological-physiological systems is not straightforward, due to their dynamic nature in which the various parts of the system interact in complex ways.
However, an understanding of these dynamics has been developed for years in the form of mathematical process simulators. We study how to incorporate this knowledge into the generative process by leveraging a biological simulator for the task of ECG classification.
Specifically, we use a system of ordinary differential equations (ODE) representing heart dynamics, and incorporate this ODE system into the optimization process of a generative adversarial network to create biologically plausible ECG training examples. 
We perform empirical evaluation and show that heart simulation knowledge during the generation process improves ECG classification.
\end{abstract}

\section{Introduction}
In the early days of AI research, it was claimed that ``the power of
an intelligent program to perform its task well depends primarily on the quantity
and quality of knowledge it has about that task'' \cite{Buchanan:1982}.
Indeed, in the last two decades it has been shown repeatedly that access to the wealth of knowledge possessed by humans significantly improves machine learning algorithms on various tasks \cite{Gabrilovich:2009:WSI,Zhang:KDD:2016}.
Knowledge has been introduced into machine-learning algorithms in a host of different ways: by enriching the features \cite{Gabrilovich:2009:WSI}, by reducing the labeling work \cite{Mintz:ACL:2009:DSR}, and by encoding conditional dependencies \cite{Wainwright:2008:GME}.
However, it is less straightforward to introduce prior knowledge about biological and physiological systems.  This is at least partially due to the fact that the complexity of biological systems stems from their dynamic nature.  Each part of the system interacts and behaves collectively to produce distinct physiologic states and functionality. To fully understand the system, one needs to model its dynamics. 
Consider the problem of detecting cardiac abnormalities from an electrocardiogram (ECG) which measures the electrical activity generated by the heart. In recent years, deep learning algorithms have been applied for the problem, yielding state-of-the-art results \cite{kachuee2018ecg, al2016deep}. 
While the performance of such algorithms is quite promising, they are purely data-driven; they do not make use of the fact that at a fundamental level the interpretation of an ECG signal involves the understanding of the electrical conduction system of the heart. 
Intuitively, it seems reasonable that knowledge about the physics of the heart's electrical and mechanical activity might improve the performance of deep learning models. 
This knowledge has been developed for years in the form of mathematical process simulators. The challenge then becomes how to incorporate this knowledge into machine learning models. In this work we tackle this challenge directly, and propose a technique which allows us to leverage a biological simulator for the task of ECG heartbeat classification.

A simulator is a model of a phenomenon or a process created by a researcher to imitate the operation of a process or system \cite{bcnn2000}. Simulation is often used to model natural systems to better understand their behavior. By changing variables in the simulation, one can make predictions about the behavior of the system over time. Simulators have been developed in a variety of fields, including chemistry, biology, and physics.
Developing such a model requires deep understanding of how the system is constructed and operates. For example, a heart simulator requires understanding of the heart mechanics and anatomy. Such a model can be used to mimic the pressure-volume relationship and simulate various conditions and pathologies.

It is common to model systems via a mathematical model, which aims to predict the system behavior given knowledge about its nature. We focus in this work on continuous simulation, where time evolves continuously. In this setting, simulation is usually performed by numerical integration of Ordinary Differential Equations (ODE), that capture the researcher's knowledge regarding the physical or biological nature of the problem. More specifically, since the ODEs cannot generally be solved in an analytic manner, numerical approximation procedures such as Runge-Kutta are used to solve the equations.
In this work, we leverage an ECG simulator proposed by \cite{mcsharry2003dynamical}; the model simulates the electrical and mechanical activity of the heart by a set of ODEs.
We explore the direct integration of these ODEs into deep learning models for the purpose of further improving the models, which have already reached state-of-the-art results for the task of ECG classification \cite{kachuee2018ecg}.

Recently, it has been shown that generating synthetic ECG signals of different arrhythmias to train deep learning models significantly improves the performance of classification models in personalized settings \cite{golany2019pgans}. 
We leverage this architecture to integrate simulator knowledge during the generation process. We hypothesize that this formulation enables generation of ECGs with better morphological characteristic of the different arrhythmias, thus improving classification. We present empirical results on gold-standard ECG datasets and show the superiority of algorithms working with simulator knowledge in the field of ECG heartbeat classification.

Our contribution in this work is threefold:
(1) We present a methodology to leverage heart simulator ODEs to improve instance generation by an adversarial network to improve a supervised classification. To the best of our knowledge, this is the first application where a physiological simulator in the form of ODEs was leveraged to improve a supervised classification task. (2) We empirically show that utilizing the synthetically generated ECG instances guided by the heart simulator ODEs significantly improves ECG heartbeat classification using deep learning techniques. 
(3) We share our code online for further research and experimentation.\footnote{\url{https://github.com/tomerGolany/sim_gan}}

The structure of our paper is as follows: 
Section \ref{sec:ecg_simulator} describes the ECG simulator represented by a system of ordinary differential equations. 
Section \ref{sec:sim_gan} introduces the framework of ECG Simulator GAN (\SimGAN{}), a GAN-based setup which learns to create synthetic data by leveraging knowledge derived from a simulator represented by a system of ODEs (in our specific case, the ECG simulator from Section \ref{sec:ecg_simulator}). Section \ref{sec:ResNet_classification} describes how the generated synthetic heartbeats from the \SimGAN{} are used to train a deep network that classifies each heartbeat according to its heartbeat class. We use a  ResNet convolutional model which was found to have superior results on the ECG gold-standard dataset \cite{kachuee2018ecg}). Finally, in Sections \ref{sec:experimental_eval}-\ref{sec:results} we present empirical evaluation comparing to state-of-the-art methods.


\section{Related Work}
\label{sec:related_work}

In imitation learning, real-world simulation has shown empirical gains.
For example, in a racing game, an agent will have access to a simulated environment, which is usually the game engine. \cite{kapoor:SIMULTECH:2019} showed that autonomous vehicle driving significantly improves after training using imitation learning in a simulated road environment.  In reinforcement learning, game simulation has been used to improve learning \cite{mnih2013playing}.

Previous work has also exploited generative models for the purposes of creating extra training data.  In the computer vision realm, \cite{shrivastava2017learning} presented a GAN model which learns to improve the realism of synthetic labeled images using unlabeled real images. The generator's input is a synthetic image (instead of random noise), and the output is a refined image. The discriminator must learn to classify images as real vs refined. They show that the improved realism enables the training of better models for the estimation of gaze and hand pose by adding the refined labeled images to the training set.  
In the ECG heartbeat classification task, \cite{golanyimproving} similarly showed that classification performance can be improved by adding synthetic ECG heartbeats generated from a standard GAN  to a training set.
Our work also generates synthetic ECG signals that can be used in training; however, unlike previous work, we leverage a simulator to improve the deep generative model's ability to generate synthetic labelled ECG heartbeats.

Classic ECG beat-level classification methods focused on extracting interval features and using prior knowledge on the ECG morphology \cite{de2004automatic}. Each ECG signal is separated into heartbeats using heartbeat detection techniques \cite{afonso1999ecg}.
For each heartbeat, features related to the heartbeat intervals and ECG morphology are calculated. The combined features are then fed into supervised machine-learning models based on linear discriminants (LDs) \cite{nasrabadi2007pattern}.
In recent years, the application of deep learning models to ECG classification has become popular. It has been applied to numerous tasks, such as cardiologist-level arrhythmia detection \cite{andrewng:arxiv:ecg} and ECG heartbeat classification \cite{guler2005ecg,prasad2003classification, kachuee2018ecg,al2016deep}.
The state-of-the-art method for ECG heartbeat-level classification \cite{kachuee2018ecg} recently showed that superior results are reached by applying a ResNet convolutional model which classifies each heartbeat class separately. 
In this work, we focus on the training of such deep learning systems trained with additional synthetic ECG heartbeats which are generated from our proposed generative model enhanced with simulator knowledge (Section \ref{sec:ResNet_classification}).


\section{The ECG Simulator}\label{sec:ecg_simulator}
In a normal resting heart, the physiological rhythm is referred to as the \emph{normal sinus rhythm} (NSR). NSR produces the prototypical pattern of the P wave, followed by the QRS complex, and finally the T wave.  To capture this pattern, \citeauthor{mcsharry2003dynamical} represented a model of the heart by a system of three ordinary differential equations (\citeyear{mcsharry2003dynamical}). 
The resulting simulator is able to generate synthetic ECG signals with realistic PQRST morphology, as well as prescribed heart rate dynamics. The simulator is parameterized by specific heart rate statistics, such as the mean and standard deviation of the heart rate, as well as frequency-domain characteristics of the heart rate variability (HRV)\cite{malik1990heart}.




The simulator is specified by three coupled ODEs, giving rise to a trajectory $(x(t), y(t), z(t))$.  The ODEs are as follows, where we have suppressed the time-dependence of $x$, $y$, and $z$ for conciseness:
\begin{align}
    \frac{dx}{dt} & = \alpha(x, y) x - \omega y \equiv f_x(x, y; \eta) \label{ode_equations_x} \\
    \frac{dy}{dt} & = \alpha(x, y) y + \omega x \equiv f_y(x, y; \eta) \label{ode_equations_y} \\
    \frac{dz}{dt} & = -\sum_{\beta \in \mathcal{B}}a_\beta\Delta\theta_\beta(x, y) e^{-\Delta\theta_\beta(x, y)^2/2b_\beta^2} - (z - z_0(t)) \notag \\
                     & \equiv f_z(x, y, z, t; \eta) \label{ode_equations_z}
\end{align}
where:


\begin{align}
    \alpha(x, y) & = 1 - \sqrt[]{x^2 + y^2} \\
    \Delta\theta_\beta(x, y) & = (\theta(x, y) - \theta_\beta)\mod 2\pi \\
    \theta(x, y) & = \text{atan2}(y, x)) \in [-\pi, \pi] \\
    \mathcal{B} & = \big\{P,Q,R,S,T\big\}
\end{align}

The so-called ``baseline wander'' has been introduced by coupling the baseline value $z_0(t)$ to the respiratory frequency $f_2$ using:
\begin{equation}
\begin{aligned}
&z_0(t) = A\sin(2\pi f_2 t)
\end{aligned}
\end{equation}
where $A=0.15$ mV.  The baseline wander is low frequency noise (0.05 - 3 Hz in stress tests)\cite{sornmo2005bioelectrical}, which is due to a variety of factors: movement during breathing, patient movement, poor contact between electrode cables and ECG recording equipment, inadequate skin preparation where the electrode has been placed, and dirty electrodes.  $\omega$ is a scalar representing the angular velocity of the trajectory as it moves around the limit cycle. 

Finally, we denote the sum total of all of the parameters of the ODEs as $\eta$; they consist of the values $\theta_\beta$, $a_\beta$, and $b_\beta$ for $\beta \in \mathcal{B} = \big\{P,Q,R,S,T\big\}$.  Explicitly, we may write
\begin{equation}
\begin{aligned}
\eta = (&\theta_P, \theta_Q, \theta_R, \theta_S, \theta_T, a_P, a_Q, a_R, a_S, a_T, \\ 
        &b_P, b_Q, b_R, b_S, b_T)
\end{aligned}
\end{equation}

The solution of the equations generates a trajectory in a three dimensional state-space with coordinates $(x, y, z)$. The generated ECG signal itself is just the third dimension, i.e. the signal $z(t)$, which is the movement of the trajectory around a limit cycle of unit radius in the $(x, y)$ plane. Each revolution around this circle corresponds to a single heartbeat (or cardiac-cycle).

\textbf{The $\eta$ parameters:} Distinct points on the ECG signal, such as the P, Q, R, S and T waves are described by events corresponding to negative and positive attractors / repellers in the $z$ direction. The wave events locations are determined by the equations parameters $\theta_P, \theta_Q, \theta_R, \theta_S, \theta_T$ which represent fixed angles along the unit circle. When the $z$ trajectory approaches one of these events, it is pushed upwards or downwards away from the limit cycle, and then it moves back toward the limit cycle. The height and length of each of the P, Q, R, S, T wave events is determined by the equations parameters $a_P, a_Q, a_R, a_S, a_T$ and  $b_P, b_Q, b_R, b_S, b_T$ respectively. We follow the $\eta$ parameters suggested by \cite{mcsharry2003dynamical} for a normal heartbeat.

\textbf{Solving the ODE system} is done numerically using \textit{Runge-Kutta methods} \cite{butcher1987numerical} with a fixed time step $\Delta{t} = \frac{1}{f_s}$ where $f_s = 360$ is the sampling frequency. Specifically, for the numerical integration of the ordinary differential equations we leverage the simplest form of the Runge Kutta family known as the Euler method~\cite{atkinson2008introduction}.

\textbf{The Euler method} is based on the finite difference approximation \cite{milne2000calculus}:
\begin{equation}
\begin{aligned}
&\frac{du}{dt}(t)\approx \frac{ u(t + \Delta{t}) - u(t)}{\Delta{t}}
\end{aligned}
\end{equation}
Given an ODE of the form $du/dt = v$, the above may be rearranged to yield the following formula:
\begin{equation}
    \label{euler_eq}
    \begin{aligned}
    &u(t + \Delta{t}) = u(t) + v(t)\Delta{t}
    \end{aligned}
\end{equation}
We construct the approximate solution as follows. We run Eq. \ref{euler_eq} iteratively for $L$ time-steps, where the $\ell^{th}$ step corresponds to $t_\ell = \ell\Delta{t}$; $L$ is the number of samples that corresponds to a single signal. This yields
\begin{equation}
    \label{euler_eq_discrete}
    \begin{aligned}
    &u_{\ell+1} = u_\ell + v_\ell\Delta{t}
    \end{aligned}
\end{equation}
Notice that approximating the values at time step $\ell + 1$ requires only the values at time step $\ell$, which are already known.  (Of course we require the initial conditions, $u_0$, as well.)

Substituting Eqs. \ref{ode_equations_x}, \ref{ode_equations_y}, \ref{ode_equations_z} into Eq. \ref{euler_eq_discrete} yields: 
\begin{equation}
    \label{euler_eq_ecg_ode}
    \begin{aligned}
    &t_\ell = \ell\Delta{t} \\
    &x_{\ell+1} = x_\ell + f_x(x_\ell, y_\ell; \eta)\Delta{t} \\
    &y_{\ell+1} = y_\ell + f_y(x_\ell, y_\ell; \eta)\Delta{t} \\
    &z_{\ell+1} = z_\ell + f_z(x_\ell, y_\ell, z_\ell, t_\ell; \eta)\Delta{t} \\
    \end{aligned}
\end{equation}



\section{\SimGAN{} Framework: \\Adding Simulators to GANs}
\label{sec:sim_gan}
We introduce the framework of ECG Simulator GAN (\SimGAN), a GAN-based setup which has been enriched with additional knowledge from the ECG simulator presented in Section \ref{sec:ecg_simulator}. 
Recall that in a regular GAN setting the generator learns to create synthetic data based on input noise. The data is then fed to a discriminator, which attempts to discriminate between synthetic and real data.
Although generating ECG signals from noise allows for the creation of samples which do not appear in the training data, it does not yield ECG heartbeats with truly realistic morphology and characteristics.  We therefore strive to incorporate the ECG simulator equations directly into the generation process.
Following the practice of \cite{golany2019pgans, golanyimproving, shrivastava2017learning}, the generated ECG signals are then used to train a deep network (Section \ref{sec:ResNet_classification}).
We empirically show (Section \ref{sec:results}) that the additional generated labeled examples significantly improve the heartbeat classification.

\subsection{General GAN Framework}\label{sec:ecg_sim_gan_framework}
We formulate the generative adversarial networks for ECG heartbeats generation as follows. An ECG signal taken from a patient in one lead is sliced to heartbeats (cardiac cycles), which are labelled $h$.  Each such heartbeat may be written as a fixed length vector $h = (h_1, \ldots, h_L)$, which represents the time evolution of the voltage values of a single heartbeat. The length $L = 216$; these 216 points represent a 600ms range sampled at 360 samples per second, where the range is from 200ms before the R-peak to 400ms after the R-peak.  Our ultimate goal is to classify the heartbeats according to their 
heartbeat type $c$. 
We denote the underlying beats distribution of a given class as a conditional probability $p(h|c)$, which reflects the distribution of heartbeat signals given that the heartbeat is taken from class $c$.

Given a set of heartbeat signals $ \{ h^{(1)}, \ldots h^{(K)} \} $ drawn from class $c$, our general GAN framework consists of a discriminator and a generator, where both are class-specific.  We denote the discriminator by $D(h ; \phi_D^c)$, where the parameters of the discriminator network are given by $\phi_D^c$, and depend on the class $c$.  The generator is denoted by $G(\mathbf{m}; \phi_G^c)$, where $\mathbf{m}$ is a random variable  with known, fixed distribution (Gaussian), and $\phi_G^c$ are the class-specific parameters.  As is standard, $G$ and $D$ play the following two-player game:
\begin{align}
    & \min_{\phi_D} L_D(\phi_D, \phi_G) \\
    & \min_{\phi_G} L_G(\phi_D, \phi_G) 
\end{align}
However, note that in our case the game is not minimax, as the discriminator and generator will have objective functions which are not equivalent, i.e. $L_D \neq -L_G$.  We now elaborate on the specific discriminator and generators we use, and their corresponding loss functions $L_D$ and $L_G$.

\subsection{The ECG Simulator Discriminator}
\textbf{Architecture} The discriminator architecture and optimization remains the same as for a conventional GAN. Given positive ECG heartbeat samples from real patients and negative ECG heartbeat samples from the generator, the objective for the discriminator is to maximize the log-probability of assigning the correct labels to both positive and negative samples.
The ECG heartbeats are fed through 6 one-dimensional convolutional layers.  All weights were initialized from a zero-centered Gaussian distribution with a standard deviation of 0.02.  After each layer we perform batch normalization and apply a Leaky ReLU activation function, where the slope of the leak is set 0.2. The final layer is a fully connected layer with sigmoid activation which classifies whether the heartbeat is from the real data or from the generator (fake).

\noindent\textbf{Loss} The discriminator's job is a typical classification problem.  Thus, we use a cross-entropy loss:
\begin{equation}
\begin{aligned}
    L_D(\phi_D, \phi_G) = & -\mathbb{E}_{h \sim p_{data}} \log D(h; \theta_D) \\
                          & -\mathbb{E}_{m \sim \mathbf{m}} \log (1-D(G(m; \theta_G); \theta_D))
\end{aligned}
\end{equation}
which is the standard GAN discriminator loss.


\subsection{The ECG Simulator Generator}
\begin{figure}[ht]
\begin{center}    
\includegraphics[width=0.48\textwidth]{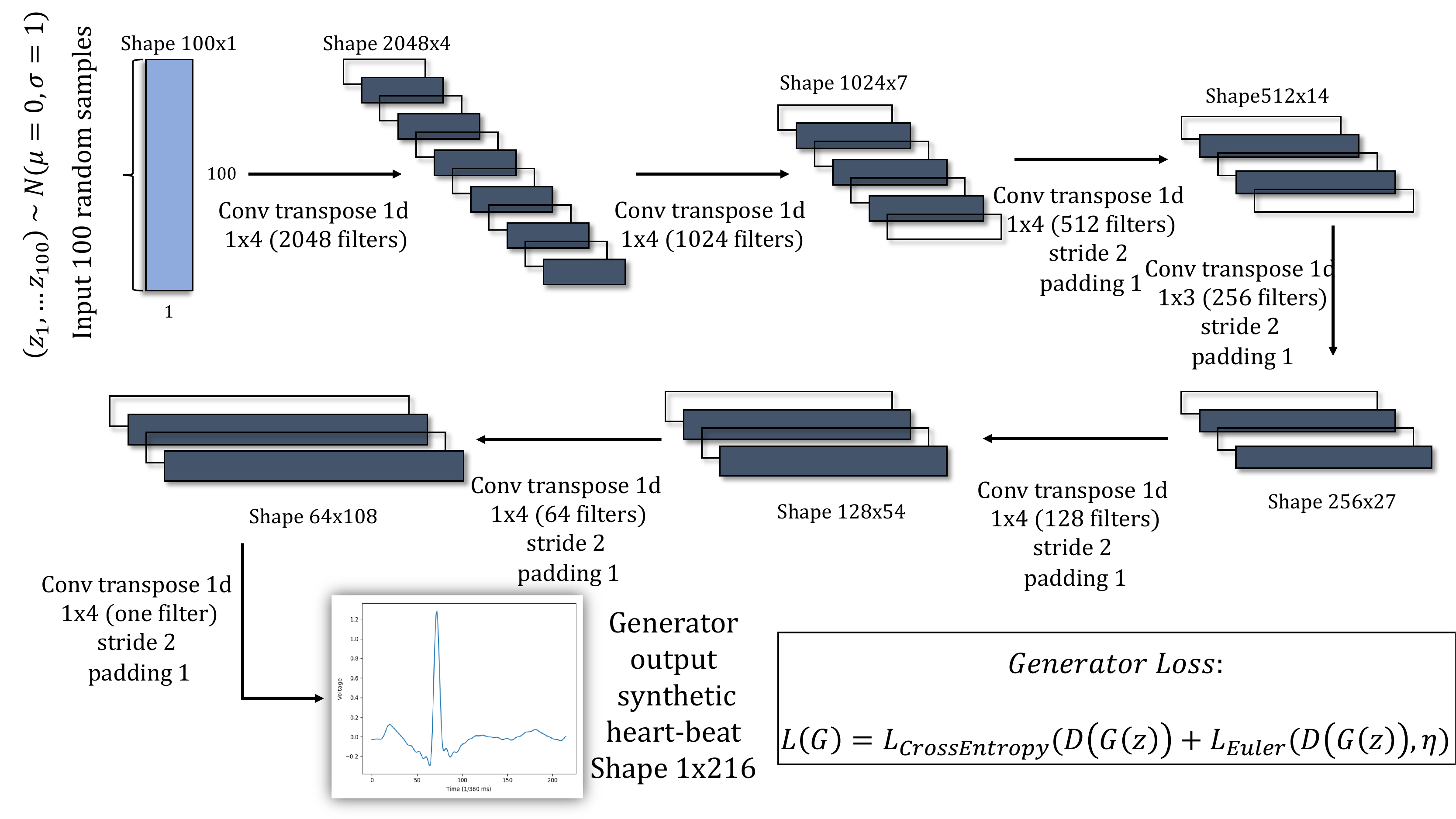}
\caption{Generator model pipeline \cite{golanyimproving}. The generator outputs a single synthetic heartbeat with characteristics similar to both the heartbeats generated from the ECG simulator as well as the real data.}
 \label{fig:generator_arch}
  \end{center}  
\end{figure}

\textbf{Architecture} The \SimGAN{} Generator architecture differs from that of the classic GAN by leveraging knowledge from the ECG simulator (Section \ref{sec:ecg_simulator}). 
Figure \ref{fig:generator_arch} describes the architecture. The generator input layer draws a length $100$ vector from a Gaussian distribution with zero mean and identity covariance: $ m \sim \mathbf{m} \equiv N(0, I)$.  This noise vector $m$ is then fed to 6 one-dimensional deconvolution layers. After each layer batch normalization \cite{ioffe2015batch} is performed followed by a ReLU activation function. 
The final one-dimensional deconvolution layer of the generator produces an output vector of size $L = 216$. The output dimensions of the generator therefore corresponds to the same dimension of one heartbeat (cardiac-cycle).

\noindent\textbf{Loss: General Considerations} The generator loss function is defined by a combination of the classical cross-entropy loss, which tries to generate cardiac-cycles that will fool the discriminator, and a Mean Square Error (MSE) function which tries to generate fake heartbeats which are morphologically similar to heartbeats from the ECG Simulator (Section \ref{sec:ecg_simulator}).  We now elaborate on each of these two loss functions.

\noindent\textbf{Cross-Entropy Loss} In contrast to the discriminator, the generator aims to minimize the log-probability that the discriminator correctly assigns negative labels to the samples generated by $G$.  That is, the goal is to optimize the generator's weights in such a way that the discriminator network will predict that the generated ECG heartbeat is real.  Formally, the cross-entropy loss function of the generator may be written
\begin{equation}
L_G^{CE}(\phi_D, \phi_G) = -\mathbb{E}_{m \sim \mathbf{m}} \log D(G(m; \theta_G); \theta_D)
\end{equation}


\noindent\textbf{Euler Loss}
In an ordinary GAN setting the generator learns to create synthetic data based on input noise which is later fed into the discriminator. We wish to leverage our world knowledge coming from the ECG simulator and use it to generalize the heartbeats produced to be more realistic. We therefore introduce the Euler Loss function.

We begin by defining a measure of how well a given heartbeat matches the ECG simulator.  If the heartbeat is given by $h$ and the simulator parameters are given by $\eta$, then the Simulator Distance is given by
\begin{equation}
\Delta_{sim}(h, \eta) = \sum_{\ell = 1}^{L-1}\left(\frac{h_{\ell+1} - h_\ell}{\Delta{t}} - f_z(x_\ell, y_\ell, h_\ell, t_\ell; \eta)\right)^2
\end{equation}
where $f_z$ is defined as in Eq. \ref{ode_equations_z}; and the trajectories for $x_\ell$ and $y_\ell$ are given by the discrete solution to the coupled ODEs in Eq. \ref{euler_eq_ecg_ode}.
%
In other words, the Simulator Distance attempts to determine how closely the generated heartbeat matches the biophysical model of a heartbeat, as specified by the set of ODEs.  In order to do so, the $z$ trajectory -- which is the heartbeat itself -- is fixed to the generated heartbeat; and the $x$ and $y$ trajectories are fixed to their ODE solutions.  The Euler Loss then measures how closely the ODE in $z$ holds.  If the generated heartbeat is exactly reproduced by the model, then the ODE in $z$ will hold exactly, leading to a Simulator Distance of $0$; the further the deviation between generated heartbeat and the biophysical model's prediction, the larger the Simulator Distance.  Note that in order for the Simulator Distance to be properly defined, we need an additional item, the initial conditions for $x$ and $y$.  These are taken to be $x_0 = -0.41, y_0 = -0.91 $ as suggested by \cite{mcsharry2003dynamical}.

Given this definition of the Simulator Distance, we define the Euler Loss as follows:
\begin{equation}
    L_G^{EUL}(\phi_G) = \mathbb{E}_{m \sim \mathbf{m}, \eta \sim p(\eta|c)} \,\, \Delta_{sim}(G(m; \theta_G), \eta)
\end{equation}
That is, the Euler Loss is the expectation of the Simulator Distance over realizations of both the noise vector input $m$ to the generator, as well as the simulator parameters $\eta$.  The distribution of the simulator parameters is approximated, for a given class $c$, as a Gaussian, where the mean and covariance matrix are computed from the set of $\eta$-vectors corresponding to that class.  (For each heartbeat in the class, one may compute the $\eta$-vector which generates a simulated heartbeat that is closest to that heartbeat.)  The covariance matrix is taken as diagonal for simplicity.  It is then straightforward to sample $\eta$.

Intuitively, then, the Euler Loss tries to constrain the output from the generator to lie close to the submanifold of signals which can produced by the simulator ODEs.  Combining the Euler Loss with the classical cross-entropy loss enables the generator to create synthetic ECG heartbeats with real morphology and characteristics which don't exist in the training set, while preserving the noise which defines the real data.

\noindent\textbf{Optimization} The final loss function of the generator is therefore the sum of the two loss functions:
\begin{equation}
L_G(\phi_D, \phi_G) = L_G^{CE}(\phi_D, \phi_G) + L_G^{EUL}(\phi_G)
\end{equation}
The optimization is performed using Adam optimizer \cite{kingma2014adam} and a learning rate of 0.0002 as suggested for training GAN models by \cite{radford2015unsupervised}. The generator's parameters are updated twice each iteration while the discriminator's parameters are updated once.

\section{Deep ECG Classification}\label{sec:ResNet_classification}
In Section \ref{sec:ecg_sim_gan_framework}, we described the \SimDCGAN{} -- a framework for generating ECG heartbeats that exhibit both a certain arrhythmia and adapt morphological characteristics from the ECG simulator (Section \ref{sec:ecg_simulator}). In this section, we describe how the generated synthetic heartbeats from the generator are used to train a deep network that classifies each heartbeat according to its %
heartbeat class.

For evaluation we use the ResNet convolutional model which was found to have superior results on the ECG gold-standard dataset \cite{kachuee2018ecg}. The network is trained over the train portion of the dataset described in Section \ref{sec:ecg_dataset}. The network architecture consists of a convolutional layer followed by five residual convolutional blocks. Each residual block contains two convolutional layers, two corresponding ReLU activations, a residual skip connection, and a pooling layer. The last residual block is followed by two fully-connected layers with 32 neurons each and a soft-max layer to predict output class probabilities. Each convolutional layer is a one-dimensional convolution through time and has 32 kernels of size 5.
Once the model's initial training has converged, we perform further training using the additional generated ECG signals from the trained generator.


\section{Experimental Evaluation}\label{sec:experimental_eval}
\subsection{ECG Dataset}\label{sec:ecg_dataset}
Our framework consists of ECG recordings taken from the MIT-BIH arrhythmia database \cite{moody2001physionet}, which is the most popular public dataset for discovering and clustering arrhythmias, and is considered the gold-standard evaluation data for ECG heartbeat classification tasks. The database contains 48 half-hour ECG records, obtained from patients studied by the BIH Arrhythmia Laboratory between 1975 and 1979. Each record contains two 30-minute ECG lead signals digitized at 360 samples per second. The database contains annotations for both heartbeat class information and timing information verified by independent experts. The database consists of a total of 109,492 heartbeats, each of which has been labelled with one of the following four classes: Ventricular Ectopic Beat, shortened to VEB or V; Supraventricular Ectopic Beat, shortened to SVEB or S; Fusion Beat, shortened to F; Normal beat, shortened to N.




\begin{figure*}
\centering     
\subfigure[SVEB heartbeat class]{\label{fig:sveb_precision_recall}\includegraphics[scale=0.25]{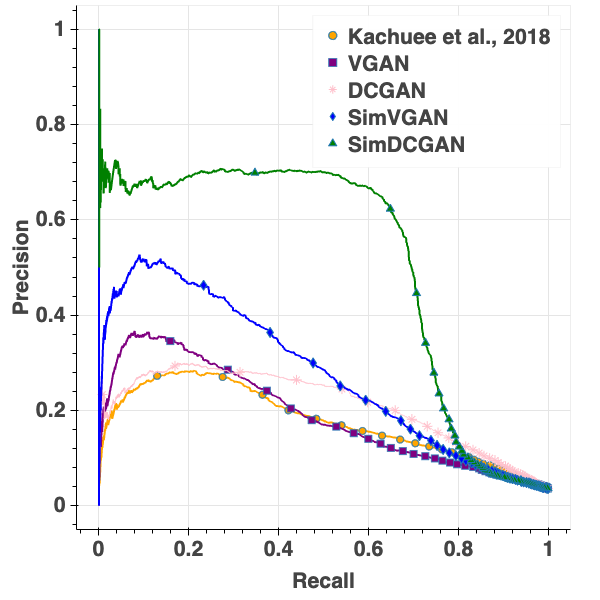}}\quad
\subfigure[Fusion heartbeat class]{\label{fig:fusion_precision_recall}\includegraphics[scale=0.25]{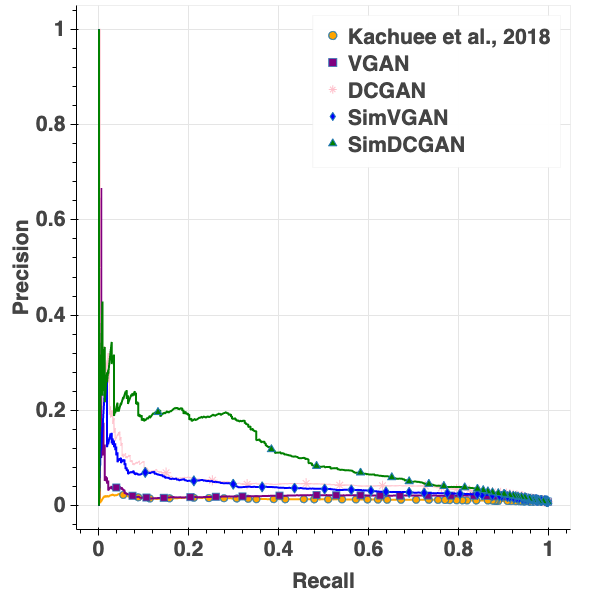}}\quad
\subfigure[VEB heartbeat class]{\label{fig:veb_precision_recall}\includegraphics[scale=0.25]{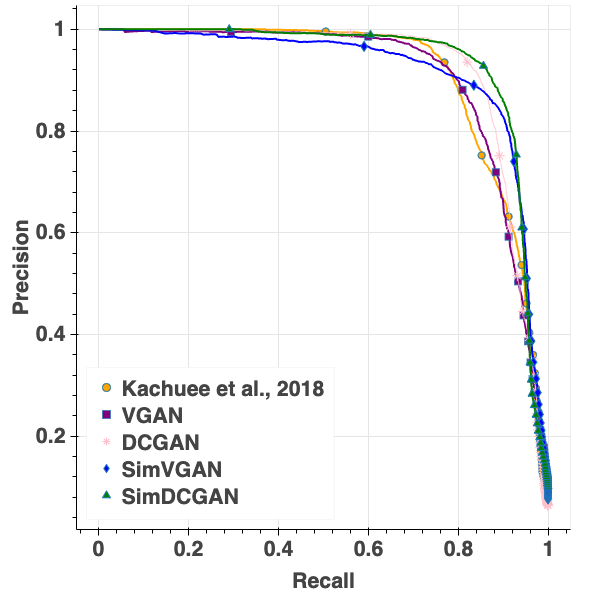}}
\caption{Precision - Recall curves of the 3 heartbeat classes evaluated on the test-set. Each subfigure shows curves for each of the four data synthesis regimes described in Section \ref{sec:experimental_eval}, as well as the state of the art method of \cite{kachuee2018ecg}.}
\end{figure*}

\subsection{Experimental Methodology}
We follow the dataset partitioning as described by \cite{de2004automatic, al2016deep, golany2019pgans}. All follow the AAMI (Association for the Advancement of Medical Instrumentation) recommendations for the ECG heartbeat classification task. This partition ensures that patient data is not mixed between the train and the test sets.  We follow the experimental methodology used in \cite{de2004automatic, al2016deep, kutlu2012feature, jiang2007block}, again as recommended by the AAMI; namely, we compute a precision-recall curve separately for each heartbeat class.  (Note that in some of the above references, the medical convention is used, namely sensitivity (=recall) vs. positive predictive value (=precision).)  For example, for the case of VEB, the precision-recall curve reports the performance in trying to distinguish V from the remaining classes S, F, and normal heartbeat.  We report similar precision-recall curves for SVEB and Fusion (no curve is reported for the normal heartbeat class, as it is not an arrhythmia and considered a simple problem with little clinical value).  As \cite{kachuee2018ecg} follows an idiosyncratic methodology different from the rest of the literature, we have re-implemented the method of \cite{kachuee2018ecg} and report results for their technique using the standard settings described above.


In what follows, we refer to the set of heartbeats from all patients in the train set as the \emph{base set}.  We present results for the following different data synthesis regimes:

\noindent\textbf{No ECG Generation}
The ResNet convolutional model which recently showed state-of-the-art results in the task of ECG heartbeat classification \cite{kachuee2018ecg} -- see Section \ref{sec:ResNet_classification} -- is trained on the base set, with no additional synthesized ECG examples. 

\noindent\textbf{ECG Generation using Ordinary GANs}
We train an ordinary GAN without any special loss terms (e.g. no Euler Loss) to generate synthetic heartbeats.  This technique has two flavors: \newline
(1) \textit{Vanilla GAN (\VGAN)} -- The ResNet convolutional model is trained on the base set with additional synthesized ECG examples from a classical GAN model \cite{NIPS2014:GAN}. \newline
(2) \textit{DCGAN} -- The ResNet convolutional model is trained on the base set with additional synthesized ECG examples from a DCGAN \cite{radford2015unsupervised} model. As DCGANs have shown superior performance on several tasks \cite{radford2015unsupervised,yeh2017semantic} we present results on generated examples from this class of generative adversarial networks.

\noindent\textbf{ECG Generation using RefineGAN} -- Generating ECG heartbeats with a DCGAN framework where the inputs to the generator are synthetic heartbeats that comes out from the ECG Simulator.
We would like to compare the contribution of integrating the ECG formulation as a loss term (The Euler loss) to a GAN, against a GAN setting where the input to the Generator are synthetic ECG heartbeats which are produced from the ECG Simulator instead of random noise. (Same as \cite{shrivastava2017learning})


\noindent\textbf{ECG Generation using ECG Simulator}
We would like to test the contribution of integrating the ECG formulation to GAN as compared to directly using generated samples from an ECG simulator. We therefore propose a baseline model which trains the ResNet convolutional model on the base set with additional synthesized ECG examples from the ECG Simulator (Section \ref{sec:ecg_simulator}).
For each of the three type of heartbeats to be classified, we calculate the values of their P, Q, R, S and T waves (Figure \ref{fig:qualitative_examples}c). 
The average values for each wave is taken and is inserted into the ECG simulator as the wave parameters  $\eta = (\theta_P, \theta_Q, \theta_R, \theta_S, \theta_T, a_P, a_Q, a_R, a_S, a_T, b_P, b_Q, b_R, b_S, b_T)$ (Section \ref{sec:ecg_simulator}). For each heartbeat generated from the ECG simulator, we add additional Gaussian noise with zero mean and identity covariance to the the wave parameters $\eta$ in order to generate multiple heartbeats from the same class.

\noindent\textbf{ECG Generation using SimGANs}
The main approach proposed in this paper. The ResNet convolutional model is trained on heartbeats from all patients from the training set with additional synthesized heartbeats from a SimGAN model. 
As with ordinary GANs, two GAN architectures with the Euler additional optimization loss are tested: \newline
(1) \textit{\SimVGAN}, using the \VGAN~architecture; \newline
(2) \textit{\SimDCGAN}, using the DCGAN architecture. \newline
The number of synthetically generated beats which are to be added to the training set is a parameter of the model, and depends on the number of samples from each class. For a heartbeat class with $n$ samples in the base set, we experimented with the following values: $0.1n, 0.3n, 0.5n, 0.8n, n, 1.5n, 2n$.

\section{Results}\label{sec:results}
\noindent\textbf{Comparison with State-Of-The-Art}
Making a fair comparison in the task of ECG heartbeat classification is somewhat challenging due to the fact that different papers use different settings. For example, previous papers presented results where heartbeats from the same subjects were shared between train and test sets \cite{jiang2006ecg}; classifiers which used more than one lead during training \cite{llamedo2012automatic}; classifiers which used RR intervals \cite{lin2014heartbeat}; and semi-supervised heartbeat classifiers which had access to unlabeled data from the test \cite{golany2019pgans}. In our setting, by contrast, we only make use of the raw ECG signals.  We therefore compare our results to techniques which are state of the art for this setting: \cite{kachuee2018ecg} and \cite{al2016deep}.

Figures \ref{fig:sveb_precision_recall} - \ref{fig:veb_precision_recall} present precision recall curves for the task of classifying the MIT-BIH ECG test-set over the heartbeat classes SVEB, Fusion and VEB respectively.  We show curves for each of the four data synthesis regimes described in Section \ref{sec:experimental_eval}, as well as the state of the art method of \cite{kachuee2018ecg}.  The SVEB and Fusion heartbeats are quite rare and only comprise 3\% and 2\% of the dataset, respectively. We see that the performance of \cite{kachuee2018ecg} on those heartbeat classes is quite low (Figures \ref{fig:sveb_precision_recall} and \ref{fig:fusion_precision_recall}, yellow curves).  Adding synthetic heartbeats from any of the generative models improves the classification performance.  However, the results of our proposed \SimDCGAN{} model are demonstrably superior to those of any other generative model: the green curves in Figures \ref{fig:sveb_precision_recall} and \ref{fig:fusion_precision_recall} are considerably higher than any of the other curves.  This phenomenon is still present in the case of the much more common VEB heartbeats, though it is less pronounced, see Figure \ref{fig:veb_precision_recall}.  In particular, the performance of \cite{kachuee2018ecg} (yellow curve) is already quite high.  Nevertheless, the \SimDCGAN{} model, again shown as the green curve, is clearly better; for example, taking two arbitrary points on the two curves, \cite{kachuee2018ecg} achieves (Recall, Precision) = (0.85, 0.75), vs. (0.88, 0.89) for \SimDCGAN{}.

\begin{table*}[!ht]
\caption{Comparison between our method and \cite{al2016deep}. The methodology of \cite{al2016deep} is to test against the entire MIT-BIH database. Re and Pr denote Recall and Precision.  Best results are shown in bold and *.}
\label{table:precision_recall_performance_compare_2}
\begin{center}
\begin{adjustbox}{max width=\textwidth}
\begin{small}
\begin{sc}
\begin{tabular}{c||c|c||c|c||c|c||c|c||c|c||c|c}
    \toprule
    \multicolumn{1}{c||}{} &
    \multicolumn{2}{c||}{\textbf{\cite{al2016deep}}} &
    \multicolumn{2}{c||}{\textbf{VGAN}} &
    \multicolumn{2}{c||}{\textbf{DCGAN}} &
    \multicolumn{2}{c||}{\textbf{SimVGAN}} &
    \multicolumn{2}{c}{\textbf{SimDCGAN}} \\
\hline
Heartbeat class & Re & Pr & Re & Pr & Re & Pr & Re & Pr & Re & Pr \\
\hline
SVEB (S) & 0.41 & 0.43 & 0.41 & 0.63 & 0.41 & 0.58 & 0.41 & 0.64 & *\textbf{0.41} & *\textbf{0.80} \\
\hline
VEB (V) & 0.91 & 0.79 & 0.91 & 0.70 & 0.91 & 0.72 & 0.91 & 0.82 & *\textbf{0.91} & *\textbf{0.84} \\
\hline
Fusion (F) & 0.60 & 0.04 & 0.60 & 0.10 & 0.60 & 0.20 & 0.60 & 0.25 & *\textbf{0.60} & *\textbf{0.40} \\

\bottomrule
\end{tabular}
\end{sc}
\end{small}
\end{adjustbox}
\end{center}
\vskip -0.2in
\end{table*}

In Table \ref{table:precision_recall_performance_compare_2} we compare our method to another state-of-the-art deep learning technique, that of \cite{al2016deep}. We perform this comparison separately, as \cite{al2016deep} report their results over the entire MIT-BIH dataset, that is, the combination of both train and test sets.  As can be seen from Table \ref{table:precision_recall_performance_compare_2}, \SimDCGAN{} far outperforms \cite{al2016deep} on SVEB and Fusion heartbeats, nearly doubling the precision for the same level of recall for SVEB, and far more than doubling the precision for the same level of recall for Fusion.  On the more common VEB heartbeats, the improvement is smaller but still present: a 6\% relative increase in precision for the same level of recall.

\noindent \textbf{Different Generative Architectures}
In Figures \ref{fig:sveb_precision_recall} - \ref{fig:veb_precision_recall} and Table \ref{table:precision_recall_performance_compare_2}, we demonstrate that learning to adapt to the ECG simulator morphology helps for different types of GAN architectures -- both \VGAN{} and DCGAN. 
Beginning with the vanilla GANs, we observe that for all types of heartbeat, results for the model trained with additional data from \SimVGAN{} outperforms the standard \VGAN: in Figures \ref{fig:sveb_precision_recall} - \ref{fig:veb_precision_recall}, the blue curve is always significantly above the purple curve.  A similar phenomena occurs also for the DCGAN model, in which \SimDCGAN{} outperforms ordinary DCGAN for all classes. Indeed, as we have already noted, \SimDCGAN{} outperforms all generative models across all classes, reaching state-of-the-art performance.  (Table \ref{table:precision_recall_performance_compare_2} supports the same conclusions numerically.)  We therefore conclude that while adding synthetic ECG heartbeats always improves ECG classification, the \SimDCGAN{} generation method yields the highest performance gains.

\begin{figure}[ht]
\centering     
\subfigure[SVEB heartbeat class]{\label{fig:simulator_sveb_precision_recall}\includegraphics[scale=0.09]{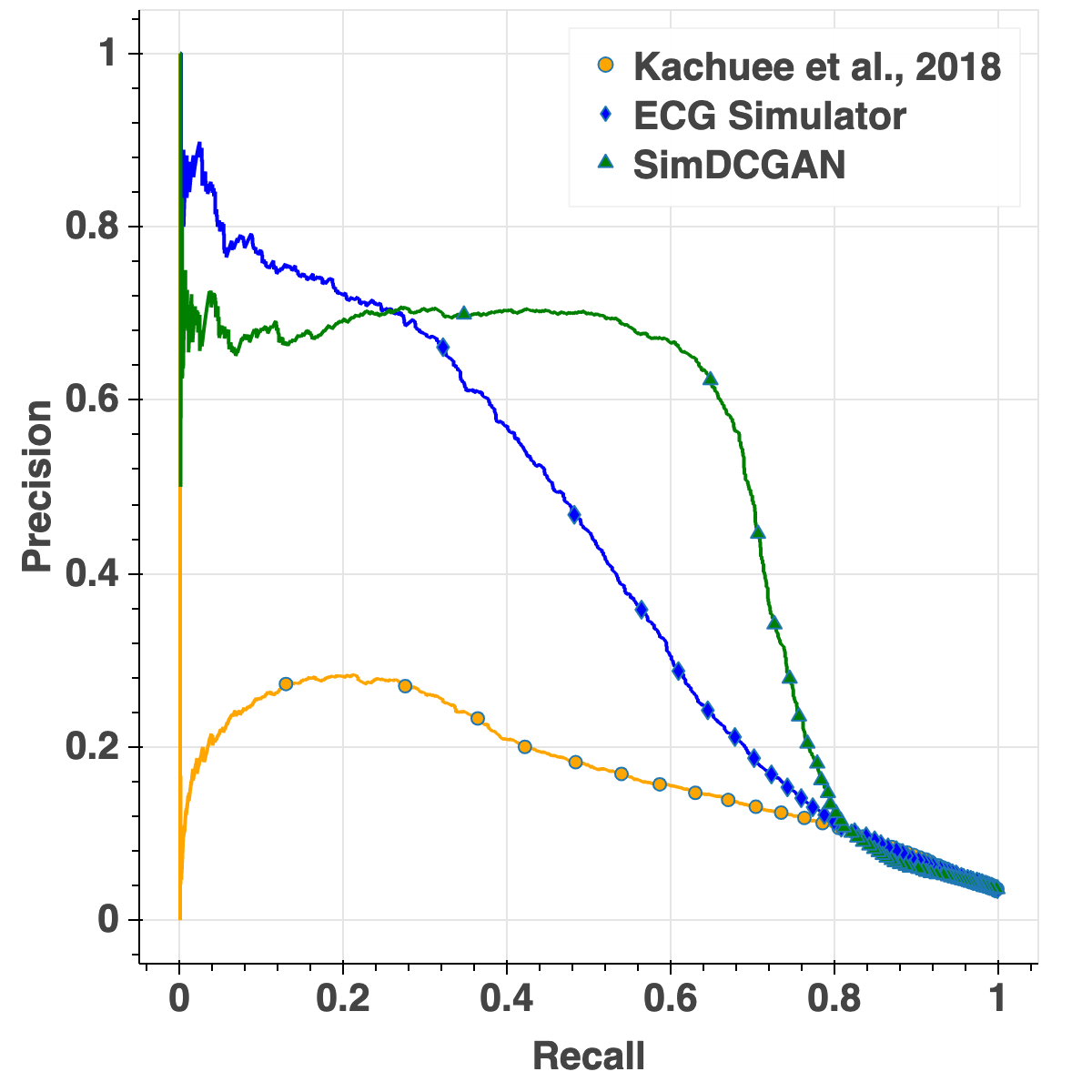}}\quad\subfigure[Fusion heartbeat class]{\label{fig:simulator_fusion_precision_recall}\includegraphics[scale=0.09]{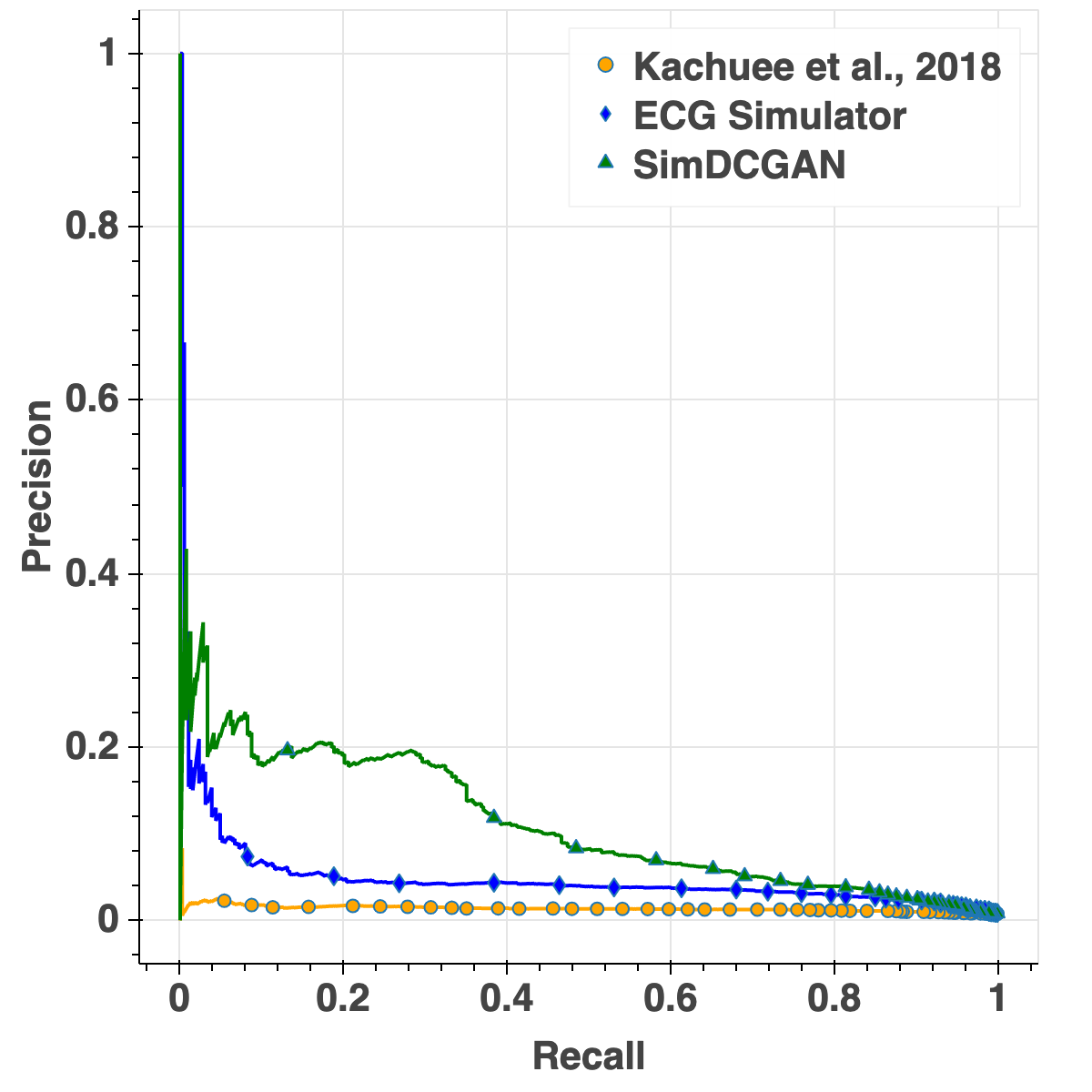}}
\vspace{-0.3cm} 
\caption{Comparison of our method against generation of synthetic heartbeats directly from the ECG Simulator. Precision-Recall curves of the two more challenging heartbeat classes.
}
\end{figure}



\noindent \textbf{The Importance of the Generative Component}
By way of an ablation study, we test whether the \SimDCGAN{} model, which relies on the Euler Loss, brings more value 
as compared to simply adding synthetic heartbeats from the ECG simulator to the training set directly. 
Figures \ref{fig:simulator_sveb_precision_recall} and \ref{fig:simulator_fusion_precision_recall} show the precision-recall curves for the two more challenging classes, SVEB and Fusion.  We observe that adding plain data from the ECG simulator outperforms the ResNet convolutional model trained only with the original training data for both heartbeat classes.  It is therefore clear that the addition of the simulator alone is of high value.
However, \SimDCGAN{} outperforms this model across both classes (with the exception of a small regime of very low recall in the case of the SVEB class).  We speculate this is due to
the fact that the data generated by the simulator does not contain the noise that exists in the real signals. It therefore does not model the real ECG data quite as well as \SimDCGAN{}.

\begin{table}[!ht]
\caption{Comparison between our method to a RefineGAN. Best results are shown in bold and *.}
\label{table:precision_recall_performance_compare_refine_gan}
\begin{center}
\begin{adjustbox}{max width=0.48\textwidth}
\begin{small}
\begin{sc}
\begin{tabular}{c||c|c||c|c||c|c}
    \toprule
    \multicolumn{1}{c||}{} &
    \multicolumn{2}{c||}{\textbf{SimDCGAN}} &
    \multicolumn{2}{c||}{\textbf{RefineGAN}} \\
\hline
Heartbeat class & Re & Pr & Re & Pr \\
\hline
SVEB (S) & *\textbf{0.45} & *\textbf{0.7} & 0.4 & 0.47 \\
\hline
VEB (V) & *\textbf{0.88} & *\textbf{0.89} & 0.85 & 0.73 \\
\bottomrule
\end{tabular}
\end{sc}
\end{small}
\end{adjustbox}
\end{center}
\end{table}
\noindent \textbf{Refining ECG Simulator output}
We study a different approach to integrate the knowledge from the ECG Simulator into a GAN framework. We adapt \cite{shrivastava2017learning} method, by inserting to the generator synthetic ECG heartbeats which comes directly from the ECG Simulator, instead of the regular setting, in which the generator receives as input random noise. In this setting the loss terms for the generator and discriminator are the conventional cross-entropy losses, without the additional Euler loss term. Table \ref{table:precision_recall_performance_compare_refine_gan} shows significant points on the precision-recall curves for the SVEB and VEB classes. \SimDCGAN{} outperforms this model across both classes.

\begin{figure}[ht]
\begin{center}    
\includegraphics[width=0.45\textwidth]{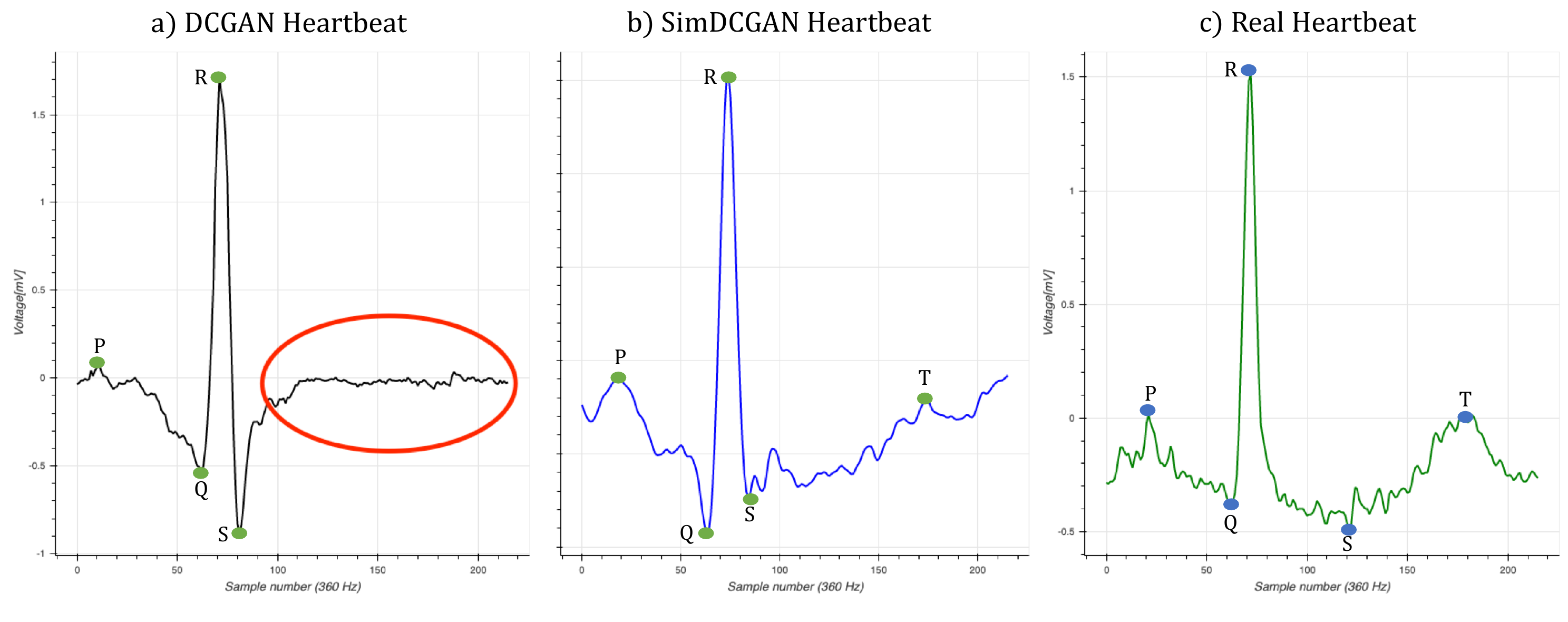}
\vspace{-0.3cm} 
\caption{\small{\textbf{a}: Heartbeat produced by DCGAN with no simulator knowledge. The red circle shows an incorrect T wave. \textbf{b}: Heartbeat produced by \SimDCGAN{}. \textbf{c}: Real patient heartbeat.}}
 \label{fig:qualitative_examples}
  \end{center}  
\end{figure}

\noindent \textbf{Qualitative Examples}
Figure \ref{fig:qualitative_examples} presents a qualitative example of a real patient ECG wave and synthetic waves produced by DCGAN and \SimDCGAN{}. Figure~\ref{fig:qualitative_examples}(c) is a typical real heartbeat taken from the dataset. The real heartbeat is slightly noisy but the PQRST waves can be observed. Figures~\ref{fig:qualitative_examples}(a) and \ref{fig:qualitative_examples}(b)  both present synthetic waves. Although both look realistic, the wave \ref{fig:qualitative_examples}(a), which shows a heartbeat produced by a DCGAN model, lacks a T wave (as shown by the red circle) and is much smoother than a real heartbeat.
Figure~\ref{fig:qualitative_examples}b shows a heartbeat generated from the \SimDCGAN{}. The heartbeat contains real ECG morphology with PQRST waves and measurement noise as well.

\vspace{-0.8cm}


\section{Conclusions}
We have presented a new technique for integrating prior knowledge in the form of a mathematical process simulator into a deep learning pipeline.  We have shown how an ODE model describing cardiac cycles can be incorporated into a GAN-based generative model of these cycles using a novel Euler Loss.  Empirically, we have demonstrated that by using the synthetic cardiac cycles generated from such a model as training data, we can improve the classification accuracy of a standard ResNet convolutional ECG heartbeat classifier, attaining state-of-the-art results.
In particular, the results attained are better than those when examples are added either (a) directly from the simulator, or (b) from ordinary GANs which do not take the simulator into account.

The approach presented here is not specific to cardiac cycles; it applies equally well to any system described mathematically by differential equations.  In the future, we will investigate the application of the approach to other systems described by ODEs, such as calcium dynamics in neuronal models \cite{de1998calcium} and chemical reactions; as well as systems described by PDEs.






\bibliography{bibliography}
\bibliographystyle{icml2020}





\end{document}